\documentclass{iopart}             
\usepackage[T1]{fontenc}
\usepackage[applemac]{inputenx}
\usepackage{graphicx}
\usepackage{url,graphics,epsfig}
\usepackage{amssymb}

\usepackage{tikz}

\newcommand*\mycirc[1]{%
  \begin{tikzpicture}[baseline=(C.base)]
    \node[draw,circle,inner sep=1pt](C) {#1};
  \end{tikzpicture}}

\begin{document}

\jl{2}
%
%
%
\def\etal{{\it et al~}}
\def\newblock{\hskip .11em plus .33em minus .07em}
%
%
%
%
%
%
\setlength{\arraycolsep}{2.5pt}             

\title[{{\it K}-shell Photoionization of B-like Oxygen (O$^{3+}$)  Ions}]{{\it K}-Shell Photoionization
	of B-like Oxygen (O$^{3+}$) Ions: Experiment and Theory}

\author{ B M McLaughlin$^{1,2}\footnote[3]{Corresponding author, E-mail: b.mclaughlin@qub.ac.uk}$, 
             J M Bizau$^{3,4}\footnote[1]{Corresponding author, E-mail: jean-marc.bizau@u-psud.fr}$,  D Cubaynes$^{3,4}$,
             M M Al Shorman$^{3}$, S Guilbaud$^{3}$,  I Sakho$^{5}$,
             C Blancard$^{6}$ and M F Gharaibeh$^{7}$ }

\address{$^{1}$Centre for Theoretical Atomic, Molecular and Optical Physics (CTAMOP),\\
			School of Mathematics and Physics, The David Bates Building, 7 College Park, Queen's University Belfast, 
			Belfast BT7 1NN, UK}

\address{$^{2}$Institute for Theoretical Atomic and Molecular Physics (ITAMP),\\
			Harvard Smithsonian Center for Astrophysics, MS-14, Cambridge, MA 02138, USA}

\address{$^{3}$Institut des Sciences Mol\'{e}culaires d'Orsay (ISMO), CNRS UMR 8214, 
			Universit\'{e} Paris-Sud, B\^{a}t. 350, F-91405 Orsay cedex, France}

\address{$^{4}$Synchrotron SOLEIL - L'Orme des Merisiers, Saint-Aubin - BP 48 91192 Gif-sur-Yvette cedex, France}

\address{$^{5}$Department of Physics, UFR of Sciences and Technologies, 
                           University Assane Seck of Ziguinchor, Ziguinchor, Senegal}
\address{$^{6}$CEA-DAM-DIF, Bruy$\grave{\rm e}$res-le-Ch\^{a}tel, F-91297 Arpajon Cedex, France}

\address{$^{7}$Department of Physics, Jordan University of Science and Technology, Irbid 22110, Jordan}


%
%

\begin{abstract}
Absolute cross sections for the {\it K}-shell photoionization of boron-like (B-like) O$^{3+}$ ions were measured by
employing the ion-photon merged-beam technique at the SOLEIL synchrotron-radiation facility in 
Saint-Aubin, France.  High-resolution spectroscopy with E/$\Delta$E $\approx$ 5000 ($\approx$ 110 meV, FWHM) 
 was achieved with photon energy from 540 eV up to 600 eV.
 Several theoretical approaches, including R-Matrix, Multi-Configuration Dirac-Fock 
 and Screening Constant by Unit Nuclear Charge were used to identify and characterize  
 the strong 1s  $\rightarrow$ 2p and the weaker  1s $\rightarrow$ 3p 
 resonances observed in the {\it K}-shell spectra of this ion.
The trend of the integrated oscillator strength and  autoionisation 
width (natural line width) of the strong $\rm 1s \rightarrow 2p$ resonances 
along the first few ions of the B-like sequence is discussed.  
\end{abstract}

%
%

\pacs{32.80.Fb, 31.15.Ar, 32.80.Hd, and 32.70.-n}

\vspace{1.0cm}
\begin{flushleft}
Short title: {\it K}-shell photoionization of B-like atomic oxygen ions\\
\submitto{\jpb: \today}
\end{flushleft}

\maketitle
%

%
%
\section{Introduction}

Single and multiply ionisation stages of C, N, O, Ne and Fe have been observed 
in the ionized outflow in the planetary nebulae NGC 4051, measured with the satellite {\it XMM-Newton} \cite{Ogle2004} 
in the soft-x-ray region. Low ionized stages of C, N and O have also been used 
in the modelling of x-ray emission from OB super-giants \cite{Cassinelli1981}.
Multiply ionization stages of O  and Fe are also seen in the {\it XMM-Newton} spectra from the Seyfert galaxy NGC 3783, including
UV imaging, x-ray and UV light curves, the 0.2 -- 10 keV x-ray continuum, the iron {\it K} - emission line, 
and high-resolution spectroscopy in the modelling of the soft x-ray warm absorber \cite{Blustin2002}.
Detailed photoionization models of the brightest cluster of star formation in the blue
compact dwarf galaxy Mrk 209 required abundances for ions of oxygen and nitrogen \cite{Diaz2007}.
O [IV] {\it K}-lines are seen in the supernova remnant Cassiopeia (Cas A) in the infrared spectra 
taken by the Spitzer Space Telescope \cite{Smith2009}.
In Seyfert galaxies based on photoionization models,  O IV comes from higher ionization states and lower density regions and 
is an accurate indicator of the power of the  active galactic nuclei (AGN) \cite{Melendez2008}.
In the present study we focus our  attention on obtaining detailed spectra on the 
triply ionized oxygen ion O$^{3+}$ (O IV) in the vicinity of its {\it K} - edge.  

Recent wavelength observations of {\it K}-transitions in atomic oxygen,
neon and magnesium and their ions with x-ray absorption lines have been made with the 
High Energy Transmission Grating (HETG) on board the {\it CHANDRA}  satellite \cite{Liao2013}. 
Strong absorption {\it K}-shell lines of atomic oxygen, in its various ionized 
forms, have  been observed by the {\it XMM-Newton} satellite 
in the interstellar medium, through x-ray spectroscopy of low-mass x-ray binaries \cite{Pinto2013}.
The {\it CHANDRA} and {\it XMM - Newton} satellite observations 
may be used for identifying the absorption features present in astrophysical
sources, such as active galactic nuclei and x-ray binaries and for assistance 
in benchmarking theoretical calculations \cite{Gatuzz2013,Gorczyca2013}.

Few experiments have been devoted to the study of {\it K}-shell photoionization on oxygen ions. 
Auger spectra of singly and doubly core-excited oxygen ions emitted in the collision of fast oxygen-ion 
beams with gas targets and foils were measured by Bruch and co-workers \cite{Bruch1979}. 
{\it K}-shell x-ray lines from inner-shell excited and ionized ions of oxygen, were observed using the Lawrence
Livermore National Laboratory EBIT. With a multi-ion model they were able to identify the 
observed {\it K}-shell transitions of oxygen ions from O$^{2+}$ to O$^{5+}$.  

Up to now, {\it K}-shell photoionization cross-section results have been obtained only for
O$^{+}$ ions by Kawatsura et al \cite{Kawatsura2002}. Measurements were made at Spring-8, using the merged-beam technique,
on relative cross sections for double photoionization spectra in the energy range of the $\rm 1s \rightarrow 2p$ resonances, 
using a limited resolving power $\sim$ 310. {\it K}-shell single and double photoionization spectra of neutral oxygen 
have also been obtained  \cite{Krause1994,Menzel1996,Stolte1997}, recently revisited with high resolution 
at the Advanced Light Source (ALS) and benchmarked against calculations using the
R-matrix with pseudo-states method \cite{Stolte2013,Oxygen2013}.

Theoretically, resonance energies and line widths for Auger 
transitions in B-like atomic ions have been calculated using a variety of methods, such 
as 1/Z perturbation theory \cite{safronova78,safronova96,safronova98,safronova99,Cornille1999}, multi-configuration 
Dirac Fock (MCDF) \cite{Chen1987,Chen1988},  the Saddle-Point-Method (SPM) with R-matrix, 
complex-coordinate rotation methods \cite{chung83,chung89,chung90,Lin2001,Lin2002}. 
Chen and Craseman \cite{Chen1987,Chen1988} calculated Auger and 
radiative decay of $1s$ vacancy states in the boron isoelectronic 
sequence using the Multi-configuration-Dirac-Fock approach (MCDF).
Sun and co-workers \cite{Sun2011} used the saddle-point method with rotation 
to calculate energy levels and Auger decay widths for the $\rm1s2s^22p^2$ 
and  $\rm 1s2s2p^3$  $^{2,4}L$  levels in B-like carbon and found suitable agreement with the 
re-calibrated spectrum of Bruch and co-workers \cite{Bruch1985} 
and the combined theoretical work and high resolution synchrotron 
measurements performed at the ALS  \cite{Schlachter2004}.
In the case of O$^{3+}$ ions, recent  saddle-point with
 rotation calculations  by Sun and co-workers \cite{Sun2013} for the energy levels and Auger and radiative decay rates 
 for the $\rm 1s2s^22p^2$ and $\rm 1s2s2p^3$  $^{2,4}L$  levels were compared to the earlier beam-foil experimental measurements 
of Bruch and co-workers \cite{Bruch1979}, the MCDF work of Chen and  Craseman  \cite{Chen1987,Chen1988}
and further extended to the higher lying $\rm 1s2p^4$ levels of other B-like  ions \cite{Sun2013b}.

State-of-the-art  {\it ab initio} calculations for Auger inner-shell 
processes were first performed on this B-like system by 
Zeng and Yuan \cite{Zeng2002} and then by Pradhan and co-workers \cite{Pradhan2003} 
using the R-matrix method \cite{rmat}. This work followed a 
similar procedure to {\it K}-shell studies on Be-like B$^+$ ions by Petrini \cite{Petrini1981}. 
 Garcia and co-workers \cite{Garcia2005}, further extended this work by using the 
R-matrix optical potential method within an intermediate-coupling scheme  \cite{Burke2011}.  
Photoionization from the ground state, along the oxygen iso-nuclear sequence was investigated, 
in the photon energy region of the {\it K}-edge. In the present study we compare our  results 
from the multi-configuration Dirac Fock (MCDF), R-matrix with pseudo-states (RMPS) approach,
and  the SCUNC semi-empirical methods  \cite{Sakho2013a,Sakho2013b} with measurements made at SOLEIL, 
 prior EBIT measurements \cite{Gu2005}, {\it XMM} and {\it CHANDRA} satellite 
 observations \cite{Blustin2002,Kaastra2005, Kallman2012, Pinto2013,Liao2013} and other theoretical results 
\cite{ Chen1987,Chen1988,Zeng2002,Pradhan2003,Garcia2005,Gu2010}.

In this paper we present detailed measurements of the  absolute {\it K}-shell  single photoionisation cross sections 
for B-like oxygen ions, in the 542--548 eV region and 594--599 eV photon energy range that were explored experimentally. 
Theoretical predictions  are made from the Screening Constant by Nuclear Unit Charge (SCUNC), 
MCDF and R-matrix with pseudo-states methods to compare with the measurements.
These calculations enable the identification and characterization 
of the very strong $\rm 1s \rightarrow 2p$ and  the weaker $\rm 1s \rightarrow 3p$ 
resonance peaks observed in the B-like oxygen spectra. 
The present investigation provides absolute values (experimental and theoretical)
for photoionization cross sections for the n=2 inner-shell resonance energies, natural line widths 
and resonance strengths, occurring for the interaction of a photon with the
$\rm 1s^22s^22p~^2P^o$ ground state and $\rm 1s^22s2p^2~^4P$ metastable state of the O$^{3+}$ ion.
Our work  would appear to be the first time that experimental measurements have been performed 
on this prototype B-like system in the photon energy region of the {\it K}-edge, and 
complements our recent studies on {\it K}-shell photoionization of atomic 
nitrogen ions \cite{Soleil2011,Soleil2013,Soleil2014} and previous 
investigations on B-like carbon, C$^+$ \cite{Schlachter2004} at the ALS, in the vicinity of the {\it K}-edge.

The layout of this paper is as follows. Section 2 briefly outlines the experimental procedure used. 
Section 3 presents the theoretical work. Section 4 presents a discussion of the
results obtained from both experiment and theory.  In section 5 
we present a discussion of the variation of the integrated oscillator strengths 
and line widths for the first three ions of the B-like sequence with increasing charge state.
Finally in section 6 conclusions are drawn from the present investigation.

\section{Experiment}\label{sec:exp}

\subsection{Ion production}

The present measurements were made using the MAIA (Multi-Analysis Ion Apparatus) 
set-up,  permanently installed on branch A of the PLEIADES beam line  \cite{Pleiades2010,Miron2013} at SOLEIL. 
The set-up and the experimental procedure have been described previously in detail \cite{Soleil2011}. 
Here we will only give a brief description of the set-up and procedure.
 A continuous oxygen ion beam is produced in a permanent magnet 
Electron Cyclotron Resonance Ion Source (ECRIS) where a micro leak (electrically driven)
is used to introduce oxygen gas at continuous rate. A 12.6 GHz radio wave is used to 
heat the plasma at a power of approximately 40W. The ions are 
extracted from the plasma by application of a constant 4 kV bias on the source.
The ion beam is selected in mass/charge ratio by a dipole magnet before being collimated 
and merged with the photon beam in the 50 cm long-interaction region. After interaction,
 the charge state of the ions is analyzed by a second dipole magnet. The parent ions are collected 
 in a Faraday cup, and the photo-ions which have lost an electron (or increase in charge state by one)
 in the interaction are counted using channel plates.
 
 \subsection{Excitation source}

The photon beam is monochromatised synchrotron radiation from the PLEIADES beam line \cite{Pleiades2010,Miron2013}. 
 In the photon energy range considered here, a permanent-magnet Apple II undulator with 80 mm 
 period is used. The light is monochromatised by a plane-grating monochromator with no entrance slit. 
 The high-flux 600 lines/mm grating was used for this work. High spectral purity is obtained by a 
 combination of a quasi-periodic design for the undulator and the use of a varied groove depth for the plane grating. 
 The photon energy is determined using a double-ionization chamber of the Samson type \cite{Samson1967}. 
 For this work, we used the $\rm 1s \rightarrow 4p \sigma$  transition in the O$_2$ gas \cite{Tanaka2008} and $\rm 3d \rightarrow$ 6p 
 transitions in Xe gas \cite{Kato2007} for calibration purposes. The photon energy is corrected for Doppler shift resulting from the velocity of the 
 oxygen ions. We note that during the experiment a misalignment in the monochromator optics 
 did not allow us to reach the optimum resolution and accuracy on the photon energy. 

\begin{figure}
\begin{center}
\includegraphics[scale=2.5,height=14.0cm,width=16.0cm]{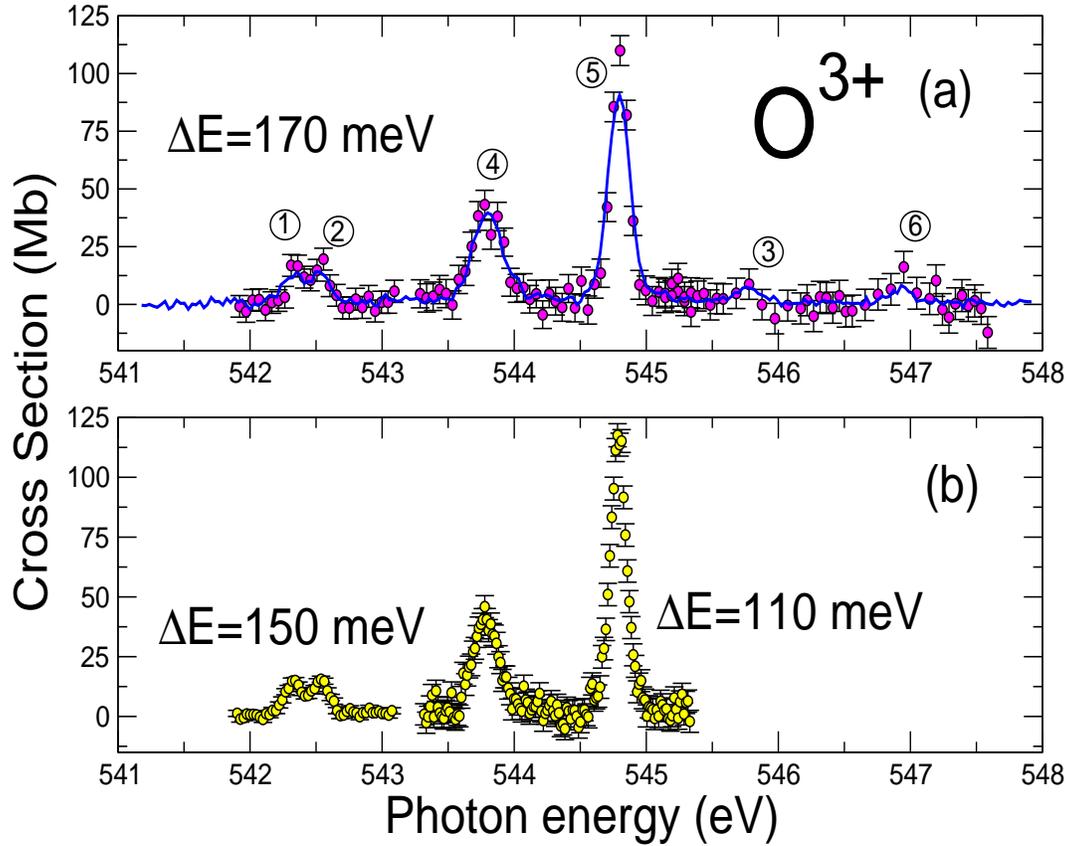}
\caption{\label{Figx1}(Colour online)  SOLEIL  experimental {\it K}-shell photoionization 
							cross section of O$^{3+}$ ions in the 541 - 548 eV photon energy range. Upper panel (a) : 
							Measured with 170 meV band-width. Solid points: the error bars give the total uncertainty; 
							blue line: normalized relative measurements.  Lower panel (b) :
							Lines 1 and 2 measured with 150 meV band-width, 
							only lines 4-5 are measured with the 110 meV band-width. 
							The error bars represent the statistical uncertainty.}
\end{center}
\end{figure}

\subsection{Experimental procedure}

The merged-beam set-up allows for the determination of absolute photoionization cross sections. 
 At a given photon energy, the cross sections $\sigma$ are obtained from,
 \begin{equation}
\sigma = \frac{S e^{2}\eta\nu q}{I J \epsilon \int^{L}_{0} \frac{dz}{\Delta x\Delta y F(z)}},
\label{eqn1}
\end{equation}
%
%
%
%
%
%
 \begin{table}
\caption{\label{expt} Typical values for the experimental parameters involved in evaluating 
			               the absolute cross section measured at  a photon energy of 544.8 eV.}
\begin{indented}
\item[]\begin{tabular}{@{}*{7}{l}}
\br
{\it S}			&70 Hz				\\
Noise		&300 Hz				\\
$\nu$		&4.2 10$^{5}$ ms$^{-1}$	\\
Photon flux	&1.3 10$^{11}$ s$^{-1}$	\\
{\it J}			&70 nA				\\
$\epsilon$	&0.48				\\
{\it F}$_{xy}$	&68					\\
\br
\end{tabular}
\end{indented}
\end{table}
\noindent
where $S$ is the counting rate of photo-ions measured with the channel-plates. 
 A chopper, placed at the exit of the photon beam line, allows to subtract from the photo-ions 
 signal the background produced by collisional-ionization processes, charge stripping on the collimator 
 slits or autoionizing decay of metastable excited states produced in the ECRIS. In (1), $q$ is the 
 charge state of the target ions, $\nu$ is the velocity of the ions in the interaction region determined from the 
 potentials applied to the ECRIS and the interaction region (see below) and $I$ is the current produced 
 by the photons on a AXUV100 IRD photodiode. The efficiency $\eta$ of the photodiode was calibrated 
 at the Physikalisch-Technische Bundesanstalt (PTB) beam line at BESSY in Berlin. $e$ is the charge 
 of the electron and $J$ is the current of incident ions measured in the Faraday cup. $\epsilon$ is the efficiency 
 of the micro-channel plates determined by comparing the counting rate produced by a low intensity 
 ion beam and the current induced by the same beam in the Faraday cup.  $\Delta x \Delta y F (z)$ is an effective 
 beam area ($z$ is the propagation axis of the two beams), where $F$ is a two-dimensional form factor 
 determined using three sets of $xy$ scanners placed at each end and in the middle of the interaction region.
  The length $L$ of the interaction region is fixed by applying -1 kV bias on a 50 cm long tube placed 
  in the interaction region, resulting in a different velocity for the photo-ions produced inside and outside the tube. 
  Typical values of the parameters involved in equation \ref{eqn1}, 
  measured at a photon energy of 544.8 eV, are given in table 1.

  The accuracy of the measured cross sections is determined by statistical fluctuations on the photo-ion and 
  background counting rates and a systematic contribution resulting from the measurement of the 
  different parameters in equation \ref{eqn1}. The latter is estimated to be 15\% and is dominated 
  by the uncertainty on the determination of the photon flux, the form factor and detector efficiency.
 To record the single-photoionization spectra, the field in the second dipole magnet is adjusted 
  to detect with the channel plates the photo-ions which have gained one charge while the 
  photon energy is scanned. Two acquisition modes have been used. One with no voltage 
  applied on the interaction tube, allowing better statistics since the whole interaction length 
  of the photon and ion beams is used. In this mode, only relative cross sections are obtained. 
  In the second mode, the voltage is applied to the tube to define the interaction length $L$, 
  allowing the determination of cross sections in absolute value.
 
 \section{Theory}\label{sec:Theory}

\subsection{SCUNC: B-like Oxygen}\label{subsec:SU_Theory}
The starting point of the Screening Constant by Unit Nuclear Charge formalism 
is the total energy of the $\left(  N\ell n\ell^{\prime}; ^{2S+1}L^{\pi}   \right ) $ 
excited states of two electron systems given by (in Rydberg units),
\begin{equation}
E  \left(  N\ell n\ell^{\prime}; ^{2S+1}L^{\pi}   \right ) 
= -Z^2  \left[  \frac{1}{N^2} 
+ \frac{1}{n^2} \left[  1 -\beta \left (  N\ell n\ell^{\prime}; ^{2S+1}L^{\pi}; Z  \right )  \right ]^2  \right ].
\end{equation}
In this equation, the principal quantum numbers $N$ and $n$ are respectively for the inner and the
outer electron of the He-like iso-electronic series. The $\beta$-parameters are screening constants by
unit nuclear charge expanded in inverse powers of $Z$ and are given by the expression,
\begin{equation}
 \beta \left( N\ell n\ell^{\prime}; ^{2S+1}L^{\pi}  \right )  = \sum_{k=1}^{~} f_k \left( \frac{1}{Z}  \right )^k 
\end{equation}
where $f_k  \left(  N\ell n\ell^{\prime}; ^{2S+1}L^{\pi}  \right )$ are parameters 
that are evaluated empirically from existing experimental measurements on resonance energies.  
In the same way one may get the Auger widths $\Gamma$ in Rydbergs (1 Rydberg = 13.60569 eV)  from the formula
\begin{equation}
\Gamma ({\rm Ry})  = Z^2  \left [  1 - \sum_{q}^{~}  f_q ~ \left (  \frac{1}{Z}   \right ) ^q   \right ] ^2 .
\end{equation}
For the B-like O$^{3+}$ ion the total energies of the $ 1sNs^xnp^y~ ^{2S+1} L$ states are given by 
\begin{eqnarray}
			&&\nonumber\\
E  \left(  1sNs^x np^y; ^{2S+1}L^{\pi}   \right )  =
			& -Z^2  \left[ 1 + \frac{x}{N^2}  \left(  1 - \sum_{k=1}^{~}f_k \left( \frac{1}{Z}  \right)^k  \right) ^2 \right]	 & \nonumber \\
			& -Z^2   \left [\frac{y}{n^2} \left( 1 -  \sum_{k=1}^{~} f^{\prime}_k \left( \frac{1}{Z}  \right)^k \right ) ^2 \right] . & \nonumber\\
			&&
\end{eqnarray}
Here $x$ and $y$ are the number of electrons in the $s$ and $p$ orbitals respectively. 
As the $\ell$-orbital quantum number is equal to zero in the $Ns^x$ orbital, 
we neglect the dependence of the  $\beta$ - parameters on $\ell$. 
This approximation permits one to simplify equation (3) for each of the $1sNs^xnp^y~ ^{2S+1} L^{\pi}$  
resonances where only one parameter  (here $f_1$) is to be evaluated empirically and similarly $f_2$ for the Auger widths. 

The Advanced Light Source experimental measurements of Schlachter and 
co-workers on {\it K}-shell photoionization  of B-like carbon ions \cite{Schlachter2004} 
were used to determine the appropriate empirical parameters $f_k$  and $f_q$ for the resonance energy and Auger widths. 
The ALS experimental data of Schlachter and co-workers on C$^+$ (Z = 6)\cite{Schlachter2004}  for the 
$\rm 1s2s^22p^2 ~^2D$, $\rm 1s2s^22p^2~^2P$, and $\rm 1s2s^22p^2~^2S$ levels are located 
at 287.93 $\pm$ 0.03, 288.40$\pm$ 0.03 and 289.90 $\pm$ 0.03, respectively, where the values are given in eV. 
Using the ground state energy of C$^+$, -1018.8467 eV \cite{NIST2012}  we obtain $f_1(\rm ^2D)$ = 1.7903 $\pm$ 0.0003,  
$f_1(\rm ^2P)$ = 1.7944 $\pm$ 0.0003  and $f_1(\rm ^2S)$ = 1.8075 $\pm$ 0.0003
for use in equation (2).  The ALS experimental values for the Auger 
widths (given in meV) on C$^+$ for these same resonances, are respectively
 105 $\pm$ 15, 59 $\pm$ 6, and 112 $\pm$ 25 and
  we find that $f_2(\rm ^2D)$ = 5.9121 $\pm$ 0.0060, $f_2(\rm ^2P)$ = 5.9341 $\pm$ 0.0033 and
 $f_2(\rm ^2S)$ = 5.9093 $\pm$ 0.0096 which are used in equation (4). The SCUNC estimates for the resonance
  energies and Auger widths for O$^{3+}$ and N$^{2+}$ using these values are given in Tables 2 -- 5.

\subsection{MCDF: B-like Oxygen}\label{subsec:MCDF_Theory}
Multi-configuration Dirac-Fock (MCDF) calculations were performed based on a full intermediate 
coupling regime in a $jj$-basis using the code developed by Bruneau  \cite{Bruneau1984}. 
Photoexcitation cross sections have been carried out for B-like atomic oxygen ions in the region of the K-edge. 
Only electric dipole transitions have been computed using length and velocity forms.
In the present study, oscillator strengths calculated using the two gauges differ by less than 8\%. 
Photoexcitation from the two levels ($\rm ^2P^{\circ}_{1/2,3/2}$) of the ground configuration $\rm 1s^22s^22p$ 
and from metastable levels ($\rm ^4P_{1/2,3/2,5/2}$) of the configuration $\rm 1s^2 2s 2p^2$
 have been calculated separately. In both cases, multiple orbitals with the same 
 quantum number have been used in order to describe the correlation and relaxation effects. 
 Wavefunctions have been calculated minimizing the Slater transition state. 
 Each electric dipole transition has been dressed by a Lorentzian profile
  with a full width half maximum (FWHM) equal to 10 meV.
  
Calculations of the $\rm ^2P^{\circ}_{1/2,3/2}$ photoexcitation cross sections 
have been performed using the following set of configurations.  The
initial configurations  used were : $\rm {\underline{1s}}^2 {\underline{2s}}^2 {\underline{np}}$ 
(with n=2, É, 5), $\rm {\underline{1s}}^2 {\underline{2s}} {\underline{2p}}{\underline{ns}}$ (with n=3, É, 5).
The final configurations were : $\rm 1s 2s^2 2p^2$, $\rm 1s 2s^2 2p np$ (with n=3, É, 5).
Such notation means that the radial functions $\rm 1s_{1/2}$ and $\rm 2p_{1/2,3/2}$ are 
 not the same for initial and final configurations. 
Similarly for the $\rm ^4P_{1/2,3/2,5/2}$ levels, photoexcitation cross sections 
   have been performed using the following set of configurations :
initial configurations : $\rm 1s^2 2s 2p np$ (with n=2, É, 4), 
$\rm 1s^2 2s 3\ell^2$ ($\ell$ =0, 1, 2), $\rm 1s^2 2s 4\ell^2$ ($\ell$=0, 1, 2), $\rm 1s^2 2s 3p 4p$, 
final configurations : $\rm 1s 2s np n^{\prime}p n^{\prime\prime}p$ (with n, n$^{\prime}$ and n$^{\prime\prime}$ =2, É, 4).

The wavefunctions have been calculated minimizing the following energy functional,
\begin{equation}
E =\frac{ \sum_{\alpha}(2J_{\alpha} + 1) E_{\alpha}}{2 \sum_{\alpha}(2J_{\alpha} + 1)}    + 
       \frac{ \sum_{\beta}(2J_{\beta} + 1) E_{\beta}}{2 \sum_{\beta}(2J_{\beta} + 1)} 
\end{equation}
where $\alpha$ and $\beta$ run over all the initial and final states, respectively. 

\begin{figure}
\begin{center}
\includegraphics[scale=2.5,height=14.0cm,width=16.0cm]{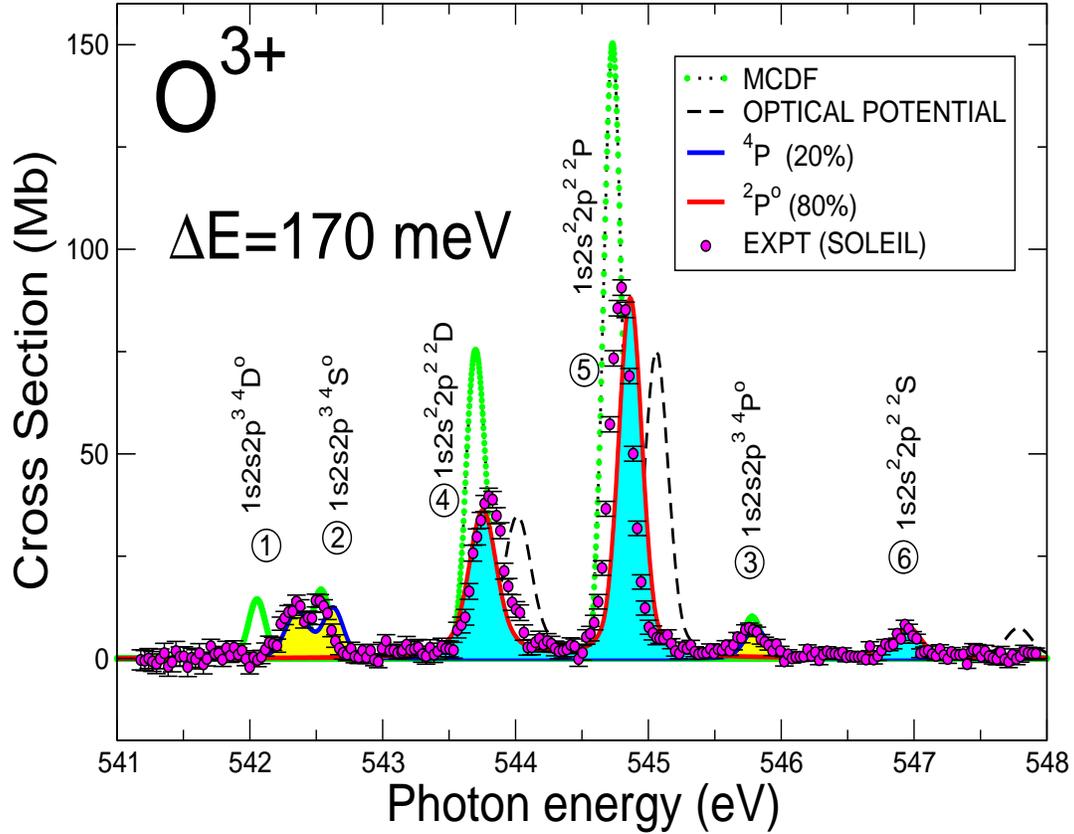}
\caption{\label{Figx2}(Colour online) Photoionization cross sections for O$^{3+}$ ions measured with a 170 meV band 
								pass in the region of $\rm 1s \rightarrow 2p$ photo-excitations. Solid points (magentta),
								experimental cross sections. The error bars give the statistical uncertainty of the experimental data. 
								The R-matrix  (RMPS, solid red line, ground state, 80\% $\rm ^2P^o$, metastable, blue line 20 \% $\rm ^4P$).
								 Dotted line with solid green circles are the present MCDF calculations (80\% $\rm ^2P^o$ and 20 \% $\rm ^4P$). 
								 The optical potential  R-matrix results (dashed black line 80\% $\rm ^2P^o$) are from 
								 the results of Garcia and co-workers  \cite{Garcia2005}. 
								 Theoretical work shown was convoluted with a Gaussian
								 profile of 170 meV FWHM and a weighting of the ground and metastable states 
								 (see text for details) to simulate the measurements. 
								 Tables \ref{reson} and \ref{reson2} 
								 gives the designation of the resonances~\mycirc{1}  - \mycirc{6}
								 in this photon energy region.}
\end{center}
\end{figure}

\subsection{R-matrix: B-like Oxygen}\label{subsec:R-Matrix_Theory}
The $R$-matrix method \cite{Burke2011}, with an efficient parallel implementation 
of the codes \cite{ballance06,McLaughlin2012,Ballance2012} was used to determine
all the cross sections presented here, for both the initial $\rm ^2P^o$ ground state 
and the  $\rm ^4$P metastable states.  Cross section calculations were carried 
out in $LS$-coupling with 390-levels  retained in the close-coupling expansion.
The Hartree-Fock $\rm 1s$, $\rm 2s$ and $\rm 2p$ tabulated orbitals of Clementi and Roetti
\cite{Clementi1974} were used with n=3 physical and n=4 pseudo
 orbitals of the O$^{3+}$ ion determined by energy optimization on the appropriate
physical and hole-shell states \cite{Berrington1997}, using the atomic structure code CIV3 \cite{Hibbert1975}.  
The n=4 pseudo-orbitals are included to account for core 
relaxation and additional correlation effects in the multi-configuration interaction wavefunctions.
All the O$^\mathrm{4+}$ residual ion states were then represented
by using multi-reference-configuration-interaction (MRCI) wave functions. The non-relativistic
$R$-matrix approach was used to calculate the energies
of the O${^\mathrm{3+}}$ bound states and the subsequent PI cross sections.
 PI cross sections  out of the $\rm 1s^22s^22p$\, $\rm ^2$P$^o$ ground state and the  
 $\rm 1s^22s2p^2$\, $\rm ^4$P metastable  state were then obtained for all total angular 
 momentum scattering symmetries that contribute.

The R-matrix with pseudo-states method (RMPS) was used 
to determine all the cross sections (in $LS$ - coupling) with
390 levels of the O$^\mathrm{4+}$ residual ion included in the 
close-coupling calculations.   
Due to the presence of metastable states in the beam, 
PI cross-section calculations were performed  
for both the $\rm 1s^22s^22p~^2P^o$ ground state and 
the  $\rm 1s^22s2p^2~^4$P meta stable state of the O$^\mathrm{3+}$ ion. 

The scattering wavefunctions were generated by
allowing three-electron promotions out of selected base
configurations of O$^\mathrm{3+}$ into the orbital set employed.
Scattering calculations were performed with twenty
continuum functions and a boundary radius of 9.4 Bohr radii.
For both the $\rm ^2$P$^o$ ground state and the  $\rm ^4P$
metastable states the outer region electron-ion collision
problem was solved (in the resonance region below and
 between all the thresholds) using a suitably chosen fine
energy mesh of 2$\times$10$^{-7}$ Rydbergs ($\approx$ 2.72 $\mu$eV)
to fully resolve all the resonance structure in the PI cross sections.  
Radiation and Auger damping were also included in the cross section calculations.  

\begin{figure}
\begin{center}
\includegraphics[scale=1.5,height=14.0cm,width=16.0cm]{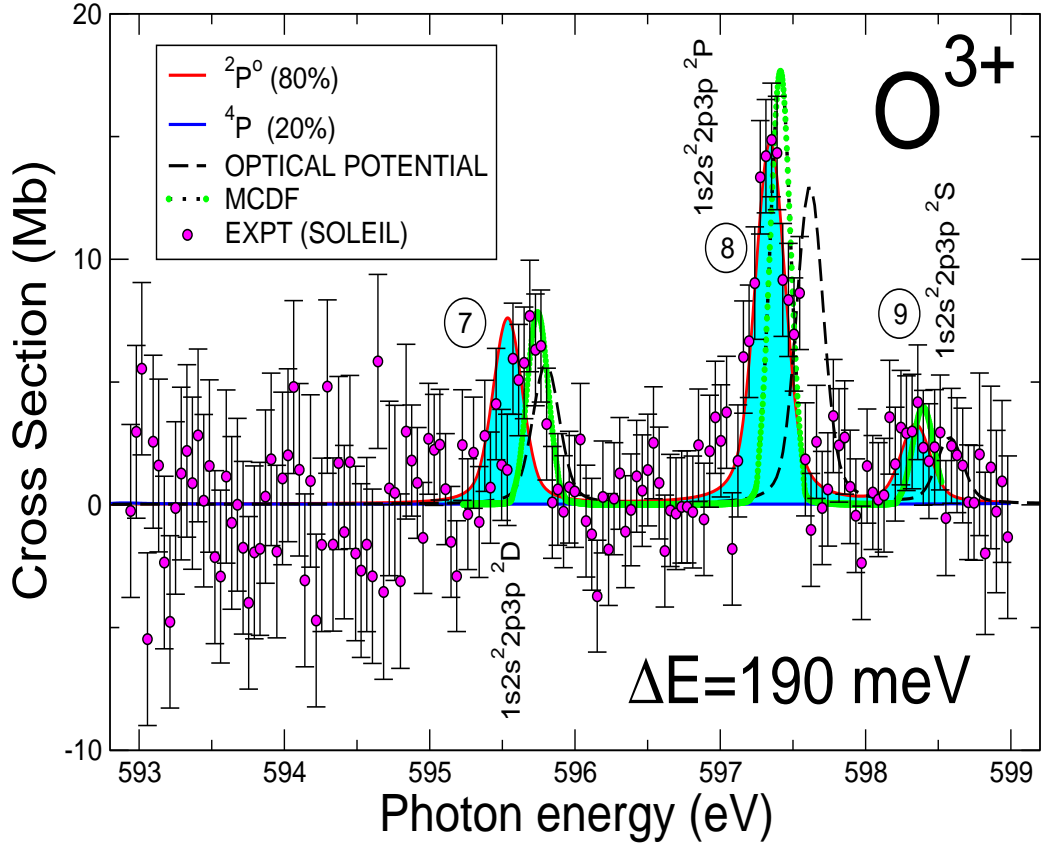}
\caption{\label{Figx3}(Colour online) Photoionization cross sections for O$^{3+}$ ions measured 
							with a 190 meV band pass in the region of the $\rm 1s \rightarrow 3p$. 
							Solid circles :  Experimental cross section measured in the relative mode, 
							normalized on the R-matrix results (see text). 
							The error bars give the statistical uncertainty of the experimental data. 
							 R-matrix  (RMPS, red line 80\% $\rm ^2P^o$,   blue line 20\% $\rm ^4P$).
							 MCDF calculations (dotted black line with green circles 80\% $\rm ^2P^o$) . 
							 The optical potential (dashed black line 80\% $\rm ^2P^o$) R-matrix results of 
							 Garcia and co-workers \cite{Garcia2005}. Theoretical work shown was convoluted with a Gaussian
							 profile of 190 meV FWHM and a weighting of the ground and metastable states 
							 (see text for details) to simulate the measurements.  
							 Table \ref{reson3} gives the designation of the 
							 resonances~\mycirc{7}  - \mycirc{9} in this photon energy region.}  
\end{center}
\end{figure}

\section{Results and Discussion}\label{sec:Results}
Figure \ref{Figx1} displays the photoionization cross sections measured at SOLEIL in the region of the $\rm 1s \rightarrow 2p$ resonances. 
In the upper panel, the continuous line shows the cross section measured in the relative mode with 170 $\pm$ 4 meV band-width (BW), 
and the open points are the ones obtained with a larger step in the absolute mode. The error bars give the total accuracy of the cross sections. 
The lower panel shows the cross sections recorded in the relative mode with improved resolution, 150 $\pm$ 25 meV BW for the region 542-543 eV, 
and 110 $\pm$ 13 meV for the region 543.3-545.3 eV. The relative measurements have been placed on an absolute scale by 
normalization to the absolute cross section by means of the area under the lines, which is independent of the experimental BW. 
The error bars on the relative measurements give the statistical uncertainty. 

{\it K}-shell photoionization contributes to the ionization balance in a more complicated way than outer-shell photoionization. 
In fact {\it K}-shell photoionization when followed by Auger decay couples three or more ionization stages instead of two 
in the usual equations of ionization equilibrium \cite{Petrini1997}. Promotion of a {\it K}-shell electron in O$^{3+}$  ions to an outer $np$-valence 
shell ($\rm 1s \rightarrow np$ transition) from the ground state produces states that can autoionise, forming an O$^{4+}$ ion and an outgoing free electron.

The 1s $\rightarrow$ 2p photo-excitation process 
on the $\rm 1s^22s^22p~^2P^o$ ground-state of B-like oxygen ion is,
$$
 h\nu + {\rm O^{3+}(1s^22s^22p~^2P^o)}  \rightarrow  {\rm O^{3+} ~ (1s2s^2 \,2p^2[^3P, ^1D, ^1S] ~^2S, ^2P, ^2D) }
 $$
 which can decay via autoionisation mainly to
  $$
{\rm  O^{4+}~ (1s^22s^2~^1S) + e^- ({\it k^2_{\ell_1}}),} \; \; {\rm or} \; \;
{\rm  O^{4+}~ (1s^22s~ np~^{1}P^{\circ}) + e^- ({\it k^2_{\ell_2}}),}
$$
where ${\it k^2_{\ell_i}}$  ($i=1,2$) represent the energies of the outgoing Auger electrons from the two different decay processes, respectively.

Three Auger  lines are expected in the spectrum, corresponding to the $\rm 1s2s^2 \,2p^2$~$\rm ^2S$, $\rm ^2P$ and $\rm ^2D$ 
resonances, due to  $\rm 1s \rightarrow 2p$ core-excited states from the ground state of the parent ion.
In the present experimental investigations, O$^{3+}$ ions are extracted from a high temperature plasma 
inside the ECRIS, where the ions are produced in all the excited states, and in particular the 
$\rm 1s^2 2s2p^2~ ^4P$ metastable state has a lifetime long enough to travel to the 
interaction region and contribute to the photoionization signal. 

For the $\rm 1s^22s2p^2~^4P$ metastable state, autoionisation processes occurring 
by the $\rm 1s  \rightarrow 2p$ photo-excitation process  are;
$$
 h\nu + {\rm O^{3+}(1s^22s2p^2~^4P)}
 $$
$$
\downarrow
$$
$$
{\rm  O^{3+} [1s2s[^{1,3}S]\,2p^3(^4S^o,^2D^o,^2P^o)]^4S^o,^4P^o,^4D^o}
$$
$$
\downarrow
$$
$$
{\rm O^{4+} (1s^22s~np~ ^3P^o) + e^- ({\it k^2_{\ell_3}}),}
$$
giving rise to three additional lines, which are 
observed in the spectra as illustrated in Figures \ref{Figx1} and \ref{Figx2},
where ${\it k^2_{\ell_3}}$  represent the energy of the outgoing Auger electron.

Figure \ref{Figx2} compares the experimental cross sections obtained with a
170 meV BW (continuous line on the top panel of Figure 1) with our theoretical 
MCDF and R-matrix results for the photon energy range of 541--548 eV. 
The R-matrix optical potential calculations (dashed line) are included for completeness.
The error bars on the experimental data give the statistical uncertainty. 
Since the relative population of $\rm ^4P$ metastable term of the O$^{3+}$ ion 
present in the parent experimental beam cannot be determined experimentally, 
theory may be used to estimate its content.  From our theoretical R- matrix studies 
on this system we find that an admixture of 80 \% ground states and 20 \% metastable states 
in the parent ion beam appears to give the closest agreement between theory and experiment. 
The same relative populations have been found in previous experimental studies 
on  B-like ions : C$^{+}$ \cite{Schlachter2004} and N$^{2+}$ \cite{Soleil2014},
performed with the same type of ion source. 

The peaks  found in  the theoretical photoionization cross sections 
spectrum were fitted to Fano profiles for overlapping resonances 
\cite{Fano1968,Shore1967,Morgan2008} as opposed to the energy derivative 
of the eigenphase sum method \cite{keith1996,keith1998,keith1999}. 
The theoretical values for the natural line widths $\Gamma$ determined from this procedure  
are presented in Tables \ref{reson} and \ref{reson2} for the strong peaks in the photon region 
540 eV to 549 eV and compared with results obtained from 
the high-resolution SOLEIL synchrotron measurements and with other methods. 
Note, in Table \ref{reson2} we have included the beam-foil values quoted by Sun {\it et al} \cite{Sun2013} 
 who reinterpreted the measurements of Bruch and co-workers \cite{Bruch1979}. 

We note the good agreement obtained by both theoretical results for the position and intensity of the lines. 
The R-matrix calculations include Auger and radiation damping, missing from the MCDF calculations 
and give lines positions and intensities in closer agreement with experiment. 
Tables \ref{reson} and \ref{reson2} summarizes our experimental, R-matrix and MCDF values for the position, width 
and strength of the observed lines, as well as the results of our SCUNC calculations with previously 
published experimental \cite{Bruch1979,Gu2005} and theoretical data \cite{Zeng2002, Pradhan2003, Garcia2005}. 
Our experimental values, including experimental band-widths, were obtained by fitting the data
with Voigt profiles to the spectra shown on Figure \ref{Figx1}. 

Concerning the excitation energy of the lines, the EBIT measurements \cite{Gu2005} give values significantly lower 
(by 0.5 eV) than our values. We note that, in the EBIT work, only the lines due to the ions in the ground states 
were observed, with an insufficient resolving power ($\sim$ 1100) to separate the individual lines. 
A comparison of the various theoretical predictions shows that our R-matrix with pseudo-states (RMPS) 
calculations give the best overall agreement with experiment, almost all energies lying within the 
experimental uncertainties. Our semi-empirical SCUNC calculations give satisfactory agreement 
with experiment, with a maximum discrepancy of 0.3 eV. Considering the work of Garcia et al \cite{Garcia2005}, 
we remark that better agreement is achieved from simple SUPERSTRUCTURE calculations 
(approximation AS1 in paper \cite{Garcia2005}) rather than by their more sophisticated R-matrix calculations. 

As in the case for  neutral oxygen \cite{Stolte2013, Oxygen2013,Gorczyca2013} there is a
discrepancy  in the position of  the strongest  $\rm 1s \rightarrow 2p$ transition in the {\it K}-shell spectrum of 
O$^{3+}$ (O IV) compared with very recent {\it Chandra} HETG observations \cite{Liao2013}
(where the $K_{\alpha}$ lines from two elements are observed at 22.6969 \AA, 546.260 eV and 22.6965 \AA, 546.270 eV).
In contrast to this the present  high resolution measurements made at SOLEIL (22.758 \AA, 544.794 eV), shows excellent agreement with previous
{\it XMM} observations (22.7769 $\pm$ 0.02 \AA,  544.340 eV and 22.75 $\pm$ 0.02 \AA, 544.985 eV) \cite{Blustin2002, Pinto2013}, 
 {\it Chandra} observations (22.74 $\pm$ 0.02 \AA,  545.225 eV  and 22.7509 $\pm$ 0.02 \AA, 544.946 eV) \cite{Kaastra2005,Kallman2012} 
 and EBIT measurements (22.741 $\pm$ 0.005 \AA,  545.201 eV) \cite{Gu2005} 
 including the current R-matrix with pseudo-states (RMPS) predictions (22.7586 \AA, 544.869 eV).  

Our measured widths are in agreement with theoretical predictions 
for the $\rm ^4S^o$, $\rm ^2D$ and $\rm ^4P^o$ lines, 
but are systematically lower compared to theory for the $\rm ^2P$ and $\rm ^2S$ lines. 
All the theoretical strengths in Tables \ref{reson} and \ref{reson2}  
have been weighted by the 80 \% and 20 \% population of the ground and metastable states, respectively.

Figure \ref{Figx3} illustrates the cross sections for the photon energy range 593--599 eV, 
in the vicinity of the $\rm 1s \rightarrow 3p$ transitions, 
which occurs from the ground state of the O$^{3+}$ ion. The measurements in this energy region were taken 
in the relative mode with a spectral resolution of 190 $\pm$ 20 meV. They are compared to the results 
of our R-matrix RMPS, MCDF and the R-matrix Optical \cite{Garcia2005}  calculations. 
The relative measurements have been normalized on the area 
of  resonance line 8 calculated with the R-matrix parallel suite of codes. 
As previously, to compare directly with experiment 
all the theoretical calculations were convoluted with a Gaussian function of 190 meV FWHM 
and weighted 80\% for the ground-state population and 20\% for the metastable state. 
In this energy region it is seen that the metastable contribution is negligible. 
Our experimental and theoretical excitation energies, widths and strengths are 
given in Table \ref{reson3}. Due to limited statistics, 
the natural width could only be obtained experimentally for the 
most intense $\rm 1s \rightarrow 3p$ transitions, 
where the strongest peak observed in this photon energy region 
is due to the $\rm 1s^22s^22p ~ ^2P^o \rightarrow 1s2s^22p3p ~^2P$ transition.  
We find an experimental value for the energy 
of this resonance to be located at 597.348 $\pm$ 0.123 eV.  
Theoretical predictions from the R-matrix with pseudo-states method (RMPS) 
give a value of 597.345 eV for the energy of this resonance and the MCDF value is 597.419 eV.
For this $\rm1s \rightarrow 3p$ resonance the SOLEIL experimental measurements 
yield a value 133  $\pm$ 71  meV  for the Auger width 
compared to the R-matrix with pseudo-states method (RMPS) of 122 meV.
SCUNC estimates give for this $\rm 1s \rightarrow 3p$ resonance 
peak an energy position of 597.629 eV and an Auger width of 37 meV.
 The R-matrix with pseudo-states method (RMPS) calculations give excitation energies 
in excellent agreement with measurements, while the predictions for this same $\rm 1s \rightarrow 3p$ resonance line 
from the  MCDF and SCUNC methods indicate the energy is at a higher energy,  respectively 0.06 eV and  0.3 eV.

In tables \ref{reson} and \ref{reson2} we have used the Heisenberg uncertainty principle  ($\Delta E \Delta t = \hbar/2$,  
i.e. $\rm \Gamma = 658.21189/ 2 \tau$, where  the natural line width $\Gamma$ is in meV
 and  $\tau$ the lifetime, is in femto-seconds)  to convert 
the MCDF and Saddle point Auger rates \cite{Chen1987,Chen1988,Sun2013} 
for the $\rm 1s \rightarrow 2p$ resonance transitions 
to widths (in meV) for comparison purposes.
%
%
%
%
%
\begin{table}
{\footnotesize
\caption{\label{reson} B-like atomic oxygen ions, quartet core-excited states. Comparison of the 
				  present experimental and theoretical results for the resonance energies $E_{\rm ph}^{\rm (res)}$ (in eV),
            			  natural line widths $\Gamma$ (in meV) and resonance strengths $\overline{\sigma}^{\rm PI}$ (in Mb eV),
           			  for the dominant core photo-excited n=2 states of the O$^{3+}$ ion, in the photon energy region 
           			  541 eV to  548 eV with previous investigations.  The experimental error in the calibrated photon 
           			  energy is estimated to be $\pm$ 70 meV.} 
 \lineup
  \begin{tabular}{ccr@{\,}c@{\,}llcl}
\br
 Resonance    & & \multicolumn{3}{c}{SOLEIL}                                       & \multicolumn{1}{c}{R-matrix} 		& \multicolumn{2}{c}{MCDF/Others}\\
  (Label)            & & \multicolumn{3}{c}{(Experiment$^{\dagger}$)}      							& \multicolumn{1}{c}{(Theory)} 		& \multicolumn{2}{c}{(Theory)}\\
 \ns
 \mr
 $\rm 1s2s[^3S]2p^3[^2D^{\circ}] \, ^4D^{\circ}$				
                                  & $E_{\rm ph}^{\rm (res)}$     	&	 	& 542.330 $\pm$ 0.082$^{\dagger}$ & 				&542.351$^{a}$ &	& 542.058$^{b}$  \\
                                  & 						 &		&~~~~~~~~~~~$\pm$ 0.013$^{*}$&					&			  &	& 			 \\
  ~ \mycirc{1}	         &           				         &		& 542.933 $\pm$ 0.12$^{\ddagger}$ &				&542.719$^{d}$ &	& 543.686$^{c}$\\
				&						&		&							&				&			  &	& 542.552$^{e}$\\
 			         &					        &		&							&				&			  &	& 542.909$^{f}$       \\
			         \\
			  	& $\Gamma$ 			       & 	   	 & 24 $\pm$ 14$^{\dagger}$ 	     	& 	         			&  \073$^{a}$ 	 &	&     	\\
 	    			&           				       &   		 &  						         &				&  \080$^{d}$ 	 &	& 38$^{c}$ 	\\ 
				&					       &		&							&				&			 &	& 96$^{e}$	\\	  
				&					       &		&							&				&			 &	& 74$^{f}$	\\	  
			         \\
				& $\overline{\sigma}^{\rm PI}$ &		& 2.49$\pm$ 0.90$^{\dagger}$		&				&\0 2.59$^{a}$	&	&2.54$^{b}$\\
 \\
  $\rm 1s2s[^1S]2p^3[^4S^{\circ}] \, ^4S^{\circ}$		 		
   				& $E_{\rm ph}^{\rm (res)}$      &   		& 542.532 $\pm$ 0.080$^{\dagger}$& 				&542.628$^{a}$&	&542.506$^{b}$  \\
                                  & 						 &		&~~~~~~~~~~~$\pm$ 0.011$^{*}$   &					&			 &	& 	 \\
  ~\mycirc{2}		&          	 		               &   		&  				 		         &				&543.203$^{d}$&	& 542.557$^{c}$ \\
   				&					      &			&							&				&			 &	& 542.209$^{e}$\\ 		  				
				&          	 		       	     &   		&  				 		  	&				&     			 &	& 543.028$^{f}$  \\
				\\
                                     & $\Gamma$                       	      &			&   23  $\pm$ 14$^{\dagger}$         	&     				& \020$^{a}$  	 &	&        \\
		 	         &                                                 &   		&  						         &				& \019$^{d}$	 &	&13$^{c}$      \\
			         &					     &			&							&				&			 &	&30$^{e}$      \\
			         &					     &			&							&				&			 &	&20$^{f}$      \\
			         
				&$\overline{\sigma}^{\rm PI}$ &		&  2.61 $\pm$ 0.99$^{\dagger}$	&				&\02.57$^{a}$	&	&3.07$^{b}$   \\
 	   		       \\
$\rm 1s2s[^3S]2p^3[^2P^{\circ}] \, ^4P^{\circ}$				
     				& $E_{\rm ph}^{\rm (res)}$    & 			& 545.730 $\pm$ 0.079$^{\dagger}$&  				&545.808$^{a}$&	&545.773$^{b}$  \\
                                  & 					    &			&~~~~~~~~~~~$\pm$ 0.010$^{*}$&					&			 &	& 			 \\
 ~\mycirc{3} 	         &           				    &   		&  							&    			 	&546.131$^{d}$&     & 547.006$^{c}$  \\
 				&					    &			&							&				&			 &	& 545.764$^{e}$ \\
				&          	 		       	     &   		&  				 		  	&				&     			 &	& 546.692$^{f}$  \\
			         \\
				& $\Gamma$			    &    		& 70  $\pm$ 23$^{\dagger}$     		&  		     		& \059$^{a}$  	 &	&  \\
 	    			&           				    &   		&  							&				& \060$^{d}$   	 &	&29$^{c}$ \\ 
				&					    &			&							&				&			 &	&76$^{e}$\\  
				&          	 		       	     &   		&  				 		  	&				&     			 &	&55$^{f}$ \\
				\\
				&$\overline{\sigma}^{\rm PI}$ &		&  1.44$\pm$ 1.10$^{\dagger}$	&				&\01.67$^{a}$	 &	&1.82$^{b}$ \\
  \br
\end{tabular}
~\\
$^{\dagger}$SOLEIL, experimental work, $^{*}$uncertainty relative to resonance line 5.\\
$^{\ddagger}$EBIT, experimental work \cite{Gu2005}.\\
$^{a}$R-matrix, RMPS present work.\\
$^{b}$MCDF, present work. \\
$^{c}$MCDF, Chen and co-workers. \cite{Chen1987,Chen1988}.\\
$^{d}$R-matrix, \cite{Zeng2002}. \\
$^{e}$SCUNC, present work.\\
$^{f}$Saddle point + complex rotation, \cite{Sun2013}.\\
}
\end{table}

%
%
%
%
%
\begin{table}
{\footnotesize
\caption{\label{reson2} B-like atomic oxygen ions, doublet core-excited states arising from the configuration 
				    $\rm 1s2s^22p^2$. Comparison of the present experimental
				    and theoretical results for the resonance energies $E_{\rm ph}^{\rm (res)}$ (in eV),
            			    natural line widths $\Gamma$ (in meV) and resonance strengths $\overline{\sigma}^{\rm PI}$ (in Mb eV),
           			   for the dominant core photo-excited n=2 states of the O$^{3+}$ ion, in the photon energy region 
           			   541 eV to  547 eV with previous investigations.  The experimental error in the calibrated photon 
           			   energy is estimated to be $\pm$ 70 meV.} 
 \lineup
  \begin{tabular}{ccr@{\,}c@{\,}llcl}
\br
 Resonance    & & \multicolumn{3}{c}{SOLEIL}                                       & \multicolumn{1}{c}{R-matrix} 		& \multicolumn{2}{c}{MCDF/Others}\\
 (Label)            & & \multicolumn{3}{c}{(Experiment$^{\dagger}$)}      & \multicolumn{1}{c}{(Theory)} 		& \multicolumn{2}{c}{(Theory)}\\
 \ns
 \mr
 $\rm 1s2s^22p^2\,[^1D] ^2$D				
  				& $E_{\rm ph}^{\rm (res)}$   & 		         & 543.801 $\pm$ 0.073$^{\dagger}$	 & 				&543.757$^{a}$        & &543.691$^{b}$  \\
                                  & 						 &		&~~~~~~~~~~~$\pm$ 0.004$^{*}$    &				&544.021$^{d}$	& &545.539$^{c}$ \\
  ~\mycirc{4} 		&           				   &   	 		& 544.614 $\pm$ 1.00$^{||}$		&				&545.465$^{e}$	& &543.809$^{g}$ \\
   		   		 &          				   &		   	&  						         &				&544.003$^{f}$     	& &544.334$^{j}$\\
   		   		 &          				   &		   	&  						         &				&     				& &	   \\
				& $\Gamma$ 			  & \;\0\0\    	& 144 $\pm$ 8$^{\dagger}$     		& 		   		&131$^{a}$       	& & \\
  	    			&          				  &   			&  							&				&140$^{d}$      	         & &\064$^{c}$ \\  
    		   		 &          				   &		   	&  						         &				& \027$^{e}$		& &151$^{g}$ \\
				&						&		&							&				& \076$^{f}$		& &155$^{j}$  \\
				\\
				&$\overline{\sigma}^{\rm PI}$ &		&  12.41 $\pm$ 2.59$^{\dagger}$ 	&				&11.85$^{a}$		& &15.32$^{b}$\\
	    			\\            
 $\rm 1s2s^22p^2\,[^3P] ^2$P				
  				& $E_{\rm ph}^{\rm (res)}$ & 			&544.794 $\pm$ 0.071$^{\dagger}$	&  				&544.869$^{a}$    	& &544.731$^{b}$ \\
                                  & 				 	&			&~~~~~~~~~~~$\pm$ 0.000$^{*}$   &				&				& & 			 \\
  ~\mycirc{5} 		&           			          &   	 		&545.201 $\pm$ 0.096$^{\ddagger}$&				&545.299$^{d}$	& & 545.068$^{c}$\\
 	    			&           				 &   			&544.340 $\pm$ 0.40$^{+}$		&				&546.909$^{e}$        & & 544.501$^{g}$\\
   		   		&          				 &		   	&544.330 $\pm$ 0.40$^{+}$		&				&545.066$^{f}$    	& & 544.945$^{h}$\\
   		  		 &           				&   	 		&545.225 $\pm$ 0.40$^{\S}$		&				&	     			& & 544.985$^{i}$ \\
     		   		&          				&		   	&544.946	$\pm$ 0.40$^{\S}$ 		&				& 		  		& & 545.379$^{j}$\\
				&     					&			&545.409 $\pm$ 1.00$^{||}$ 		&				&				& &	\\
				\\
				& $\Gamma$			 & \;\0\0\0    	&  35 $\pm$ 4$^{\dagger}$    		&  		     		& \075$^{a}$   		& &    \\
 	    			&          				 &   			&  							&				& \067$^{d}$   	         & & \028$^{c}$   \\              
   		   		 &          				 &		   	&  						         &				& \014$^{e}$		& &\078$^{g}$	 \\
				&					&			&							&				& \076$^{f}$		& &\092$^{j}$ 	 \\			\\   
				&$\overline{\sigma}^{\rm PI}$&			& 19.68$\pm$2.59$^{\dagger}$ 	&				&17.45$^{a}$		& &27.90$^{b}$\\
 	   		      \\
 $\rm 1s2s^22p^2\,[^1S] ^2$S				
   				& $E_{\rm ph}^{\rm (res)}$ & 		         & 546.908 $\pm$ 0.085$^{\dagger}$	& 				&546.954$^{a}$    	& &548.783$^{b}$ \\
                                 	& 					&			&~~~~~~~~~~~$\pm$ 0.016$^{*}$&					&				& & 	\\
  ~\mycirc{6} 		&           				&   			&547.914 $\pm$ 1.00$^{||}$ 		&				&547.276$^{d}$	& & 547.066$^{c}$ \\	
   		  		 &           				&   	 		& 					 		&				&552.022$^{e}$     	& & 546.711$^{g}$ \\
   		  		 &           				&   	 		&  						         &				&547.791$^{f}$     	& & 547.624$^{j}$ \\
				 \\
 				& $\Gamma$			& \;\0\0\    		&  26 $\pm$ 22$^{\dagger}$	 	&    				& 130$^{a}$  		& &   	\\
	    			&          				&   			&  							&				& 125$^{d}$    		& & \058$^{c}$\\  
   		   		&          				&		   	&  						         &				& \022$^{e}$   		& &162$^{g}$	\\
				&					&			&							&				&127$^{f}$		& &\091$^{j}$\\  
				\\ 
				&$\overline{\sigma}^{\rm PI}$&			& 3.73 $\pm$ 1.67$^{\dagger}$	&				&1.87$^{a}$		& &3.05$^{b}$\\
  \br
\end{tabular}
~\\
$^{\dagger}$SOLEIL, experimental work, $^{*}$uncertainty relative to resonance line 5.\\
$^{\ddagger}$EBIT, experimental work \cite{Gu2005}.\\
$^{||}$BEAM-FOIL, experimental work, assignments are from Sun and co-workers \cite{Sun2013}.\\
$^{+}${\it XMM}, \cite{Blustin2002,Pinto2013} and $^{\S}${\it CHANDRA} observations, \cite{Kaastra2005,Kallman2012}.\\
$^{a}$R-matrix, RMPS present work\\
$^{b}$MCDF, present work, $^{c}$Chen and co-workers. \cite{Chen1987,Chen1988}.\\
$^{d}$R-matrix,  \cite{Zeng2002}, $^{e}$R-matrix,  \cite{Pradhan2003}, $^{f}$R-matrix, optical potential \cite{Garcia2005}.\\
$^{g}$SCUNC, present work. \\
$^{h}$Flexible Atomic Code (FAC), \cite{Gu2010}.\\
$^{i}$HFR,Garcia and co-workers,\cite{Garcia2005}.\\
$^{j}$Saddle-point + complex rotation, \cite{Sun2013}.\\
}
\end{table}

%
%
%
%
%
\begin{table}
\caption{\label{reson3} B-like atomic oxygen ions, doublet core-excited states arising from the configuration 
				    $\rm 1s2s^22p3p$. Comparison of the present experimental
				    and theoretical results for the resonance energies $E_{\rm ph}^{\rm (res)}$ (in eV),
            			    natural line widths $\Gamma$ (in meV)
           			   for these photo-excited states of the O$^{3+}$ ion, in the photon energy region 
           			   594 eV to  599 eV with previous investigations.  The experimental error in the calibrated photon 
           			   energy is estimated to be $\pm$ 110 meV.} 
 \lineup
  \begin{tabular}{ccr@{\,}c@{\,}llcl}
\br
 Resonance    & & \multicolumn{3}{c}{SOLEIL}                                       & \multicolumn{1}{c}{R-matrix} 		& \multicolumn{2}{c}{MCDF/Others}\\
 (Label)            & & \multicolumn{3}{c}{(Experiment$^{\dagger}$)}      & \multicolumn{1}{c}{(Theory)} 		& \multicolumn{2}{c}{(Theory)}\\
 \ns
 \mr
 $\rm 1s2s^22p3p\, ^2$D					
  				& $E_{\rm ph}^{\rm (res)}$   & 		         &  595.671 $\pm$ 0.134$^{\dagger}$& 		&595.534$^{a}$        		& &595.749$^{b}$  \\
                                 	& 					&			&							&		&595.806$^{c}$		& &595.977$^{d}$\\
     ~\mycirc{7}				 \\
				& $\Gamma$ 			  & \;\0\0\    	&   --    						& 		&\090$^{a}$       		& &   \\
                                 	& 					&			&							&		&\066$^{c}$			&  &\091$^{d}$ \\
				    \\  
	    			\\            
 $\rm 1s2s^22p3p\, ^2P$				
  				& $E_{\rm ph}^{\rm (res)}$ & 			&  597.348 $\pm$ 0.123$^{\dagger}$&  		&597.352$^{a}$    		& &597.419$^{b}$    \\
                                 	& 					&			&							&		&597.616$^{c}$		& &597.629$^{d}$\\
   ~\mycirc{8}		&          				 &		   	&  						         &		&			 		& & \\
				\\
				& $\Gamma$			 & \;\0\0\0    	& 133 $\pm$ 71$^{\dagger}$	 	&  		 &122$^{a}$   			& & 	\\
   		   		 &          				 &		   	&  						         &		&\075$^{c}$			& & 37$^{d}$	 \\
				\\
 	   		      \\
 $\rm 1s2s^22p3p\, ^2$S				
   				& $E_{\rm ph}^{\rm (res)}$ & 		         & 598.362 $\pm$ 0.175$^{\dagger}$	& 		&598.347$^{a}$    		& &598.382$^{b}$ \\
                			& 					&			&							&		&598.585$^{c}$		& &598.309$^{d}$\\
     ~\mycirc{9}		&          				 &		   	&  						         &		&			 		& &  	 \\
     \\
 				& $\Gamma$			& \;\0\0\    		&  	--	 					&    		&\098$^{a}$  			&  &   \\
     		   		 &          				 &		   	&  						         &		&\087$^{c}$			& &100$^{d}$ \\
  \br
\end{tabular}
~\\
$^{\dagger}$SOLEIL, experimental work.\\
$^{a}$R-matrix, RMPS present work.\\
$^{b}$MCDF, present work. \\
$^{c}$R-matrix, optical potential \cite{Garcia2005}.\\
$^{d}$SCUNC, present work. \\
\end{table}

%
%
%
%
%
\begin{table}
{\footnotesize
\caption{\label{osci-all} B-like  ions, doublet core-excited states arising from the configuration 
				    $\rm 1s2s^22p^2$. Comparison of the present experimental
				    and theoretical results for the integrated oscillator strengths $f$ and the
            			    natural line widths $\Gamma$ (in meV) 
           			   for the dominant core photo-excited n=2 states of the first three B-like ions with previous investigations.} 
 \lineup
  \begin{tabular}{ccr@{\,}c@{\,}llcl}
\br
 Resonance    & & \multicolumn{3}{c}{SOLEIL/ALS}                              & \multicolumn{1}{c}{R-matrix} 		& \multicolumn{2}{c}{Others}\\
 (Label)            & & \multicolumn{3}{c}{(Experiment)}      & \multicolumn{1}{c}{(Theory)} 		& \multicolumn{2}{c}{(Theory)}\\
 \ns
 \mr
 $\rm 1s2s^22p^2\,[^1D] ^2$D		&		&			&							&				&				& &	\\
  \\			
C$^{+}$	  		& $f$   				  & 		         & 0.05 $\pm$ 0.015$^{\ddagger}$ 	& 				&0.069$^a$		& &  \\
				& $\Gamma$ 			  & \;\0\0\    	& 105 $\pm$ 15$^{\ddagger}$ 		& 		   		&102$^a$			& & \\
				&					 &			&							&				&103$^b$			& & \\
				\\ 		
N$^{2+}$	  		& $f$   				  & 		         & 0.115 $\pm$ 0.02$^{\dagger}$	& 				&0.119$^{a}$		& &  \\
				& $\Gamma$ 			  & \;\0\0\    	& 122 $\pm$ 19$^{\dagger}$     	& 		   		&122$^{a}$       	& &123$^{f}$\\
  	    			&          				  &   			&  							&				&109$^{e}$      	         & & \\  
 \\			
O$^{3+}$	  		& $f$   				  & 		         & 0.142 $\pm$ 0.03$^{\dagger}$	& 				&0.135$^a$		& &0.135$^c$ \\
				&					 &			&							&				&0.137$^d$ 		& &0.174$^{h}$\\
				\\
				& $\Gamma$ 			  & 		    	& 144 $\pm$ 8$^{\dagger}$  		& 		   		&131$^{a}$       	& &151$^{f}$  \\
  	    			&          				  &   			&  							&				&140$^{d}$      	         & &155$^{g}$ \\  
    		   		&          				  &		   	&  						         &				& \076$^{e}$		& & \\
 $\rm 1s2s^22p^2\,[^3P] ^2$P	&			  &			&							&				&				& &	\\
  \\			
C$^{+}$	  		& $f$   				  & 		         & 0.10 $\pm$ 0.03$^{\ddagger}$ 	& 				&0.135$^a$		& &  \\
\\
				& $\Gamma$ 			  & \;\0\0\    	& \059 $\pm$ 6$^{\ddagger}$ 		&		   		&\056$^a$		& &	 \\
				&					 &			&							&				&\062$^b$		& & \\
				\\ 		
N$^{2+}$	  		& $f$   				  & 		         & 0.191 $\pm$ 0.03$^{\dagger}$	& 				&0.220$^a$		& &  \\
 				& $\Gamma$			 & \;\0\0\0    	& 58 $\pm$  7$^{\dagger}$  		&  		     		& \0\062$^{a}$   	& & \066$^{f}$  \\
 	    			&          				 &   			&  							&				&  \0\043$^{e}$   	& &   \\              
 	    			&          				 &   			&  							&				& 		 	         & &   \\              
		
  O$^{3+}$		& $f$ 				& 			& 0.224 $\pm$ 0.03$^{\dagger}$	&  				&0.200$^a$    		& &0.243$^c$    \\
				&					 &			&							&				&0.253$^d$		& &0.318$^{h}$ \\
				\\
				& $\Gamma$			 & \;\0\0\0  		&  35 $\pm$ 4$^{\dagger}$     		&  		     		& \075$^{a}$   		& &\078$^{f}$  \\
 	    			&          				 &   			&  							&				& \067$^{d}$   	         & &\092$^{g}$ \\              
   		   		 &          				 &		   	&  						         &				& \076$^{f}$		& & \\
				&					&			&							&				& 				& & \\			   
 $\rm 1s2s^22p^2\,[^1S] ^2$S	&			&			&							&				& 				& &\\
  \\			
C$^{+}$	  		& $f$   				  & 		         & 0.008 $\pm$ 0.002$^{\ddagger}$ 	& 				&0.014$^a$		& & \\	
				& $\Gamma$ 			  & \;\0\0\    	& 112 $\pm$ 25$^{\ddagger}$ 		&		   		&105$^a$			& & \\
				&					 &			&							&				&\093$^b$		& & \\
\\ 		
N$^{2+}$	  		& $f$   				  & 		         &  0.014 $\pm$ 0.005$^{\dagger}$	& 				& 0.017$^a$		 & &  \\
 				& $\Gamma$			& \;\0\0\    		& 120  $\pm$ 60$^{\dagger}$		&    				& 106$^{a}$  		 & & 132$^{f}$  \\
	    			&          				&   			&  							&				& \094$^{e}$    		& &  \\  
 \\			
   O$^{3+}$		& $f$ 				& 		          & 0.043$\pm$ 0.02$^{\dagger}$	& 				&0.029$^a$   		& &0.035$^c$ \\
  				&					 &			&							&				&0.027$^d$		& &0.030$^{h}$\\
				\\
				& $\Gamma$			& 	   		&  26 $\pm$ 22$^{\dagger}$	 	&    				& 130$^{a}$  		& &162$^{f}$\\
	    			&          				&   			&  							&				& 125$^{d}$    		& &\091$^{g}$\\  
   		   		&          				&		   	&  						         &				& 127$^{e}$		& & \\
 \br
\end{tabular}
\\
$^{\dagger}$SOLEIL, experimental work.\\
$^{\ddagger}$ALS, experimental work \cite{Schlachter2004}.\\
$^{a}$R-matrix, RMPS, 390 levels present work, $^{b}$R-matrix, RMPS, 135 levels \cite{Schlachter2004}. \\
$^{c}$CIV3. structure calculations, \cite{Zeng2002}.\\
$^{d}$R-matrix,  \cite{Zeng2002}, $^{e}$R-matrix, optical potential \cite{Garcia2005}.\\
$^{f}$SCUNC, present work \\
$^{g}$Saddle-point + complex rotation, \cite{Sun2013}.\\
$^{h}$MCDF, present work.
}
\end{table}

\begin{figure}
\begin{center}
\includegraphics[scale=1.5,height=14.0cm,width=16.0cm]{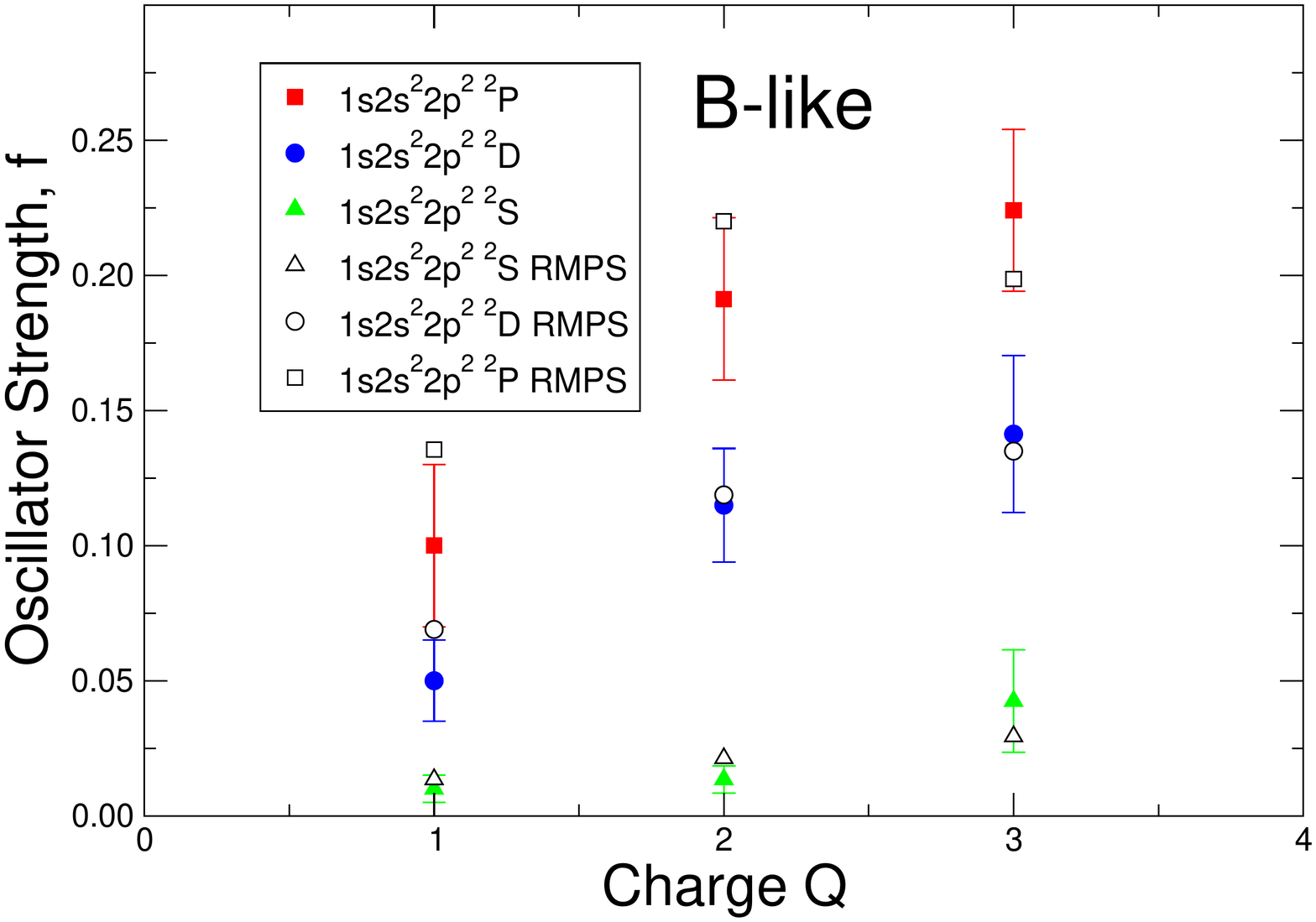}
\caption{\label{Figx4}(Colour online) Oscillator strengths $f$ for the $\rm 1s2s^22p^2~^2P, ^2D, ^2S$ Auger states of
                              				the first three ions in the B-like sequence versus increasing  charge $Q$. 
							The Auger states, $\rm 1s2s^22p^2~^2S$ (green solid triangles),
							$\rm 1s2s^22p^2~^2D$  (blue solid circles), $\rm 1s2s^22p^2~^2P$  (red solid squares), are
							 from experimental studies on C$^+$\cite{Schlachter2004}, N$^{2+}$  \cite{Soleil2014} 
							and the present O$^{3+}$ investigation.  Theoretical results, 
							R - matrix with pseudo-states (RMPS, open triangles $\rm 1s2s^22p^2~^2S$, 
							open squares $\rm 1s2s^22p^2~^2P$ , open circles $\rm 1s2s^22p^2~^2D$, ),
							see table \ref{osci-all} for numerical values.}
\end{center}
\end{figure}

\section{Auger widths and $f$-values for B-like ions; C$^{+}$, N$^{2+}$ and O$^{3+}$ ions}\label{sec:Isoelectronic}
To gain some physical insight into how a photon interacts with ions along the 
B-like sequence, we draw comparison with previous and more recent experimental 
and theoretical work on the B-like ions; C$^{+}$  \cite{Schlachter2004},  N$^{2+}$  \cite{Soleil2014}, 
O$^{3+}$ \cite{Sun2013} and the current investigation on O$^{3+}$ in the {\it K}-shell region.
Results for the cross sections and resonance parameters from the R-matrix with pseudo-states  (RMPS)
approximation provide accurate results which are used here for comparison purposes across all three ions.
In order to have a consistent comparison with the earlier ALS experimental data made
on B-like carbon ion, C$^{+}$  \cite{Schlachter2004},  RMPS calculations were 
performed on the C$^{+}$ B-like ion using the same 390-level approximation presently used for O$^{3+}$ 
and previously for N$^{2+}$ \cite{Soleil2014}  ions. Resonance parameters were extracted
and a consistent comparison made between the experimental and theoretical work for all three ions 
of the B-like sequence, not available currently with other theoretical approaches.

\begin{figure}
\begin{center}
\includegraphics[scale=1.5,height=14.0cm,width=16.0cm]{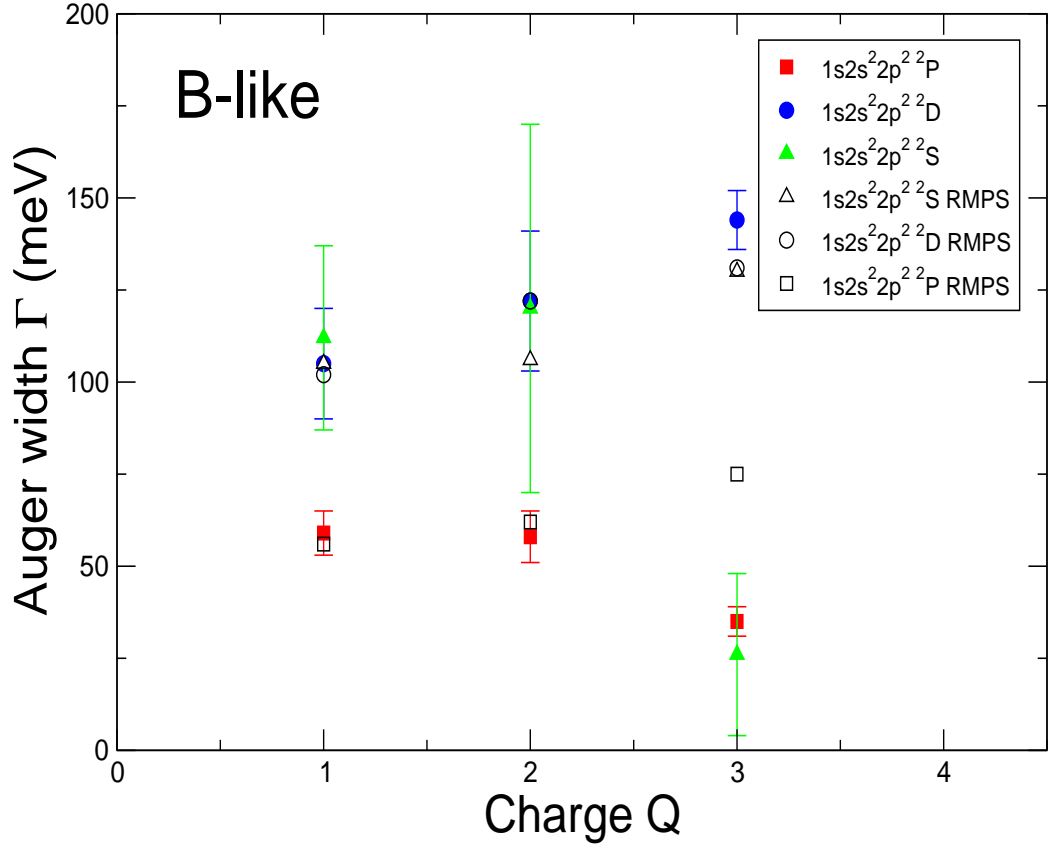}
\caption{\label{Figx5}(Colour online) Autoionization resonance widths $\Gamma$ (meV) for the $\rm 1s2s^22p^2~^2P, ^2D, ^2S$ Auger states of
                              				the first three ions in the B-like sequence versus increasing charge $Q$. 
							The Auger states, $\rm 1s2s^22p^2~^2S$ (green solid triangles),
							$\rm 1s2s^22p^2~^2D$  (blue solid circles), $\rm 1s2s^22p^2~^2P$  (red solid squares), are
							 from experimental studies on C$^+$\cite{Schlachter2004},  N$^{2+}$  \cite{Soleil2014} 
							and the present O$^{3+}$ investigation.  Theoretical results, 
							R - matrix with pseudo-states (RMPS, open triangles $\rm 1s2s^22p^2~^2S$, 
							open squares $\rm 1s2s^22p^2~^2P$ , open circles $\rm 1s2s^22p^2~^2D$, ),
							see table \ref{osci-all} for numerical values.}
\end{center}
\end{figure}

The indirect photoionization cross section for each  resonance can be derived 
from the calculated oscillator strengths,
 where the integrated oscillator strengths $f$ of the spectra, may be calculated using,
 \begin{eqnarray}
 \sigma(E)   &=& 2 \pi^2 \alpha  a_0^2 ~ R_{\infty}     \frac{df}{dE} \nonumber  \\
                     &= & 1.097618 \times 10 ^{-16}                 \frac{df}{dE}  \quad       {\rm eV ~ cm^2}
 \end{eqnarray}
where $\alpha$ is the fine structure constant, $a_0$ is the Bohr radius, $R_{\infty}$ is the Rydberg constant 
and $df/dE$ is the differential oscillator strength per unit energy  \cite{Cowan1981,Fano1968}.   
Rearranging and integrating we have the expression for the oscillator strength $f$ namely,
\begin{equation}
 f =9.11 \times 10^{-3} \int_{E_1}^{E_2} \sigma (E) {\rm d} E \label{osc}
 \end{equation}
where $\sigma (E)$ is the photoionization cross section in Mb (1 Mb = 1.0 $\times$ 10$^{-18}$ cm$^2$), 
$E$ is the photon energy in eV, and $E_1$ and $E_2$ are the appropriate energy limits of the range for which the 
photoionization cross section is to be calculated. 

Experimental and theoretical $f$-values were determined for the strong $\rm 1s \rightarrow 2p$ core-excited resonances, 
resulting from 100\%  of the parent ion in  the ground states
for the first three members of the B-like sequence. The integrated oscillator strengths $f$ (equation \ref{osc}) 
are presented in figure \ref{Figx4} and numerical values in table \ref{osci-all} compared with other theoretical methods.  
We see from figure \ref{Figx4} and table \ref{osci-all} that the area under the peaks of 
the resonances for the R-matrix with pseudo-states (RMPS) 390-level approximation 
are in suitable agreement with the available experimental studies for all three ions. 
Similarly in figure  \ref{Figx5}, (for all three  parent ions) and 
table  \ref{osci-all} we present experimental and theoretical values
 (determined using a variety of methods)  for the resonant Auger widths $\Gamma$ (meV) of these 
same states.  While there is satisfactory agreement between the 
R-matrix theoretical results and experiment for the integrated oscillator 
strength $f$ along the isoelectronic sequence (as illustrated in figure \ref{Figx4}) from figure \ref{Figx5},
we see differences for the O$^{3+}$ parent ion  between theory and experiment for the resonance 
Auger widths of both the $\rm 1s2s^22p^2~^2P$ and $\rm 1s2s^22p^2~^2S$ core-excited
states.  The R-matrix with pseudo-states calculations (RMPS) and Saddle point  calculations 
predict a monotonically increasing of the width for the three core-excited states, 
whereas in the case of the O$^{3+}$ ion the present SOLEIL experiment show 
strong decreasing for the both the $\rm ^2P$ and $\rm ^2S$ resonance 
Auger states.  Furthermore, from table \ref{osci-all} this  behaviour 
 is not supported by other theoretical methods for the same resonance widths of these Auger states, 
 as consistency between the different state-of-the art  theoretical  approaches (where electron correlation 
 is satisfactory included)  is found.  The difference between theory and experiment for these two resonance 
 is as yet unexplained so further independent investigation would be desirable to resolve this issue.
 
\section{Conclusions}\label{sec:Conclusions}
{\it K}-shell photoionization cross sections for B-like oxygen ions, O$^{3+}$, have been determined
using state-of-the-art experimental and theoretical methods.
High-resolution spectroscopy was able to be achieved with E/$\Delta$E = 5000, 
covering the energy range 540--630 eV.  Several strong resonance peaks are found in the 
cross sections in the energy region  542--548 eV and 593--599 eV.
These resonance peaks are identified as the 1s $\rightarrow$ 2p and  1s $\rightarrow$ 3p transitions  
 in the O$^{3+}$ {\it  K}-shell spectrum and assigned spectroscopically with
 their resonance parameters tabulated in Tables \ref{reson}, \ref{reson2} and \ref{reson3}.
 For the observed resonance peaks, suitable agreement is found between the present theoretical and
experimental results both on the photon-energy scale and on the absolute
 cross-section scale for this prototype B-like system. 
 A comparison between theory and experiment 
 for the integrated oscillator strengths $f$ and Auger autoionization 
 widths $\Gamma$, for the strong $\rm 1s \rightarrow 2p$ resonances
  of the first few members of  the B-like isoelectronic 
 sequence highlight differences, particularly in the present experimental 
 studies on the O$^{3+}$ ion for the Auger widths of the 
 $\rm 1s2s^22p^2~^2P$ and $\rm 1s2s^22p^2~^2S$ core-excited states. 
 These differences are as yet unexplained and would require further independent investigations.

The strength of the present study is the high resolution of  the spectra along with 
 theoretical predictions made using the state-of-the-art MCDF, R-matrix with pseudo-states methods
 and predictions from a semi-empirical approach.  Cross section results from earlier 
 R-matrix investigations (Garcia and co-workers  \cite{Garcia2005}, Pradhan and co-workers \cite{Pradhan2003})
 were restricted to the ground-state of this ion.   
 The present results have been compared with high resolution 
 experimental measurements made at the SOLEIL 
 synchrotron radiation facility and with other theoretical methods so
 would be suitable to be incorporated into astrophysical modelling codes like 
 CLOUDY \cite{Ferland1998,Ferland2003}, XSTAR \cite{Kallman2001} 
 and AtomDB \cite{Foster2012} used to numerically
simulate the thermal and ionization structure of ionized astrophysical nebulae. 

\ack
The experimental measurements were performed on the 
PLEIADES beam line, at the SOLEIL Synchrotron 
radiation facility in Saint-Aubin, France.  The authors would like to thank the SOLEIL staff and, 
in particular C Miron the local contact of the PLEIADES beam line during the experiment 
for their helpful assistance.  M F Gharaibeh acknowledges 
 funding from the Scientific Research Support Fund, Jordan, for supporting a research
 visit to SOLEIL, under contract number Bas/2/02/2010.
B M McLaughlin acknowledges support from the US National Science Foundation through a grant to ITAMP
at the Harvard-Smithsonian Center for Astrophysics, the RTRA network {\it Triangle de la Physique} 
and a visiting research fellowship from Queen's University Belfast. 
We thank John C Raymond and Randall K Smith at the Harvard-Smithsonian Center 
for Astrophysics for discussions on the astrophysical applications.
The computational work was carried out at the National Energy Research Scientific
Computing Center in Oakland, CA, USA, the  Kraken XT5 facility at the National Institute 
for Computational Science (NICS) in Knoxville, TN, USA
and at the High Performance Computing Center Stuttgart (HLRS) 
of the University of Stuttgart, Stuttgart, Germany. 
Stefan Andersson from Cray Research is acknowledged for his advice and assistance with 
the implementation of the parallel R-matrix codes on the Cray-XE6 at HLRS.
The Kraken XT5 facility is a resource of the Extreme Science and Engineering Discovery Environment (XSEDE), 
which is supported by National Science Foundation grant number OCI-1053575. 
This research also used resources of the Oak Ridge Leadership Computing Facility 
at the Oak Ridge National Laboratory, which is supported by the Office of Science 
of the U.S. Department of Energy under Contract No. DE-AC05-00OR22725.
%
%
%
%
\section*{References}
\bibliographystyle{iopart-num}
\bibliography{o3plus}

\end{document}